# Information Flow and Computation in the Maxwell Demon Problem

**Roger D. Jones[a,b], Sven G. Redsun[a], Roger E. Frye[a]**

[a]CommodiCast, Inc. and Complexica, Inc.
CommodiCast Building
552 Agua Fria
Santa Fe, NM 87501
Roger.Jones@Complexica.com
www.complexica.com
www.commodicast.com

[b]Applied Theoretical Physics Division
Los Alamos National Laboratory
Los Alamos, NM 87545 USA

November 28, 2003

## Abstract

**In this paper we examine the Maxwell Demon problem from an information theoretic and computational point-of-view. In particular we calculate the required capacity of a communication channel that transports information to and from the Demon. Equivalently, this is the number of bits required to store the information on a computer tape. We show that, in a simple model for the Maxwell Demon, the entropy of the universe increases by at least an amount $\eta \approx 0.8399955201358$ bits/particle in going from unsorted to sorted particles and by an amount $\eta^* \approx 2.373138220832$ bits/particle in going from one sorted state to another sorted state.**

**Key Words: Maxwell Demon, sorting, information flow, computation**

## 1. Introduction

In this paper we address an old question that originated with Maxwell nearly 140 years ago and can be stated simply as "Can measurement and intelligence overcome the tyranny of the Second Law of Thermodynamics?" Maxwell imagined an intelligent microscopic "Demon" who could separate fast particles from slow particles in a box, thus creating a temperature gradient without increasing the entropy of the universe – unless the Demon itself increased its personal entropy in the process. Here, we examine the Maxwell Demon[1][2][3] problem from a modern information theoretic and a computer science point-of-view, rather than from a physical point-of-view. Specifically, we

Complexica Report 031128

calculate the required capacity of a communication channel that transports information to and from the Demon. We imagine that our Demon is an idealized computer composed of a set of tapes that can be written and erased. The entropy increase of the universe is measured as the sum of the entropy decrease of the sorted particles plus the number of bits in a minimal computer program that sorts the particles.

This paper complements a treatment[4] of the Maxwell Demon problem that takes a purely physical point-of-view. There, the entropy increase of the universe is calculated from a traditional thermodynamic approach in which the Demon is represented by an ensemble of purely physical computational engines. The ensemble approach and the computational approach yield the same results. The equivalence of the two approaches was first pointed out by Zurek who also pointed out that entropy can be partitioned, as we have done here, into a piece that is purely statistical and a piece that is purely computational.[5] The statistical piece of the entropy is just the missing information in the problem and the computational piece is the minimal number of bits needed to describe the known information. The advantage of the approach taken here is that it can be easily generalized to nonphysical systems. It allows us, for example, to treat financial markets as thermodynamic systems.[6]

The approach leads to very specific quantitative results. We show that the minimum entropy increase of the universe due to sorting particles from an initially random state is

$$\delta S = \eta N \text{ (measured in bits)} \qquad (1.1)$$

where $N$ is the total number of particles in the system and

$$\eta = \frac{\pi^2}{12\ln(2)} - \frac{\ln(2)}{2} \qquad (1.2)$$

$$\approx 0.8399955201358 \text{ bits}$$

The minimum entropy increase if the particles are initially sorted, but sorted incorrectly (each particle is initially on the wrong side of the box) is

$$\delta S^* = \eta^* N \text{ (measured in bits)} \qquad (1.3)$$

where

$$\eta^* = \frac{\pi^2}{6\ln(2)}$$

$$= 2\eta + \ln(2) \qquad (1.4)$$

$$\approx 2.373138220832 \text{ bits}$$

2Complexica Report 031128

calculate the required capacity of a communication channel that transports information to and from the Demon. We imagine that our Demon is an idealized computer composed of a set of tapes that can be written and erased. The entropy increase of the universe is measured as the sum of the entropy decrease of the sorted particles plus the number of bits in a minimal computer program that sorts the particles.

This paper complements a treatment[4] of the Maxwell Demon problem that takes a purely physical point-of-view. There, the entropy increase of the universe is calculated from a traditional thermodynamic approach in which the Demon is represented by an ensemble of purely physical computational engines. The ensemble approach and the computational approach yield the same results. The equivalence of the two approaches was first pointed out by Zurek who also pointed out that entropy can be partitioned, as we have done here, into a piece that is purely statistical and a piece that is purely computational.[5] The statistical piece of the entropy is just the missing information in the problem and the computational piece is the minimal number of bits needed to describe the known information. The advantage of the approach taken here is that it can be easily generalized to nonphysical systems. It allows us, for example, to treat financial markets as thermodynamic systems.[6]

The approach leads to very specific quantitative results. We show that the minimum entropy increase of the universe due to sorting particles from an initially random state is

$$\delta S = \eta N \text{ (measured in bits)} \qquad (1.1)$$

where $N$ is the total number of particles in the system and

$$\eta = \frac{\pi^2}{12\ln(2)} - \frac{\ln(2)}{2} \qquad (1.2)$$

$$\approx 0.8399955201358 \text{ bits}$$

The minimum entropy increase if the particles are initially sorted, but sorted incorrectly (each particle is initially on the wrong side of the box) is

$$\delta S^* = \eta^* N \text{ (measured in bits)} \qquad (1.3)$$

where

$$\eta^* = \frac{\pi^2}{6\ln(2)}$$

$$= 2\eta + \ln(2) \qquad (1.4)$$

$$\approx 2.373138220832 \text{ bits}$$

Although we will not discuss it, the value of $\eta$ in Eq. (1.4) also appears in one dimensional chaotic time series[7] as the Liapunov exponent for the Gauss map.

## *2. The Maxwell Demon*

Maxwell wished to address the question of the role of intelligence in the flow of entropy. The Demon was an intelligent agent that sat at a trapdoor separating a box into two sides. Particles inhabited both sides of the box. The Demon observed the particles and allowed fast particles to enter into one side of the box and slow particles to enter into the other side of the box. Particles that tried to escape from their proper sides were thwarted by a closed trapdoor. The entropy of the particles was thus decreased and a temperature gradient, capable of producing useful work, was created. The intelligent Demon seemed to violate the Second Law of Thermodynamics. Either the Second Law had to be abandoned or the entropy of the Demon had to increase to compensate for the decrease in entropy of the particles.

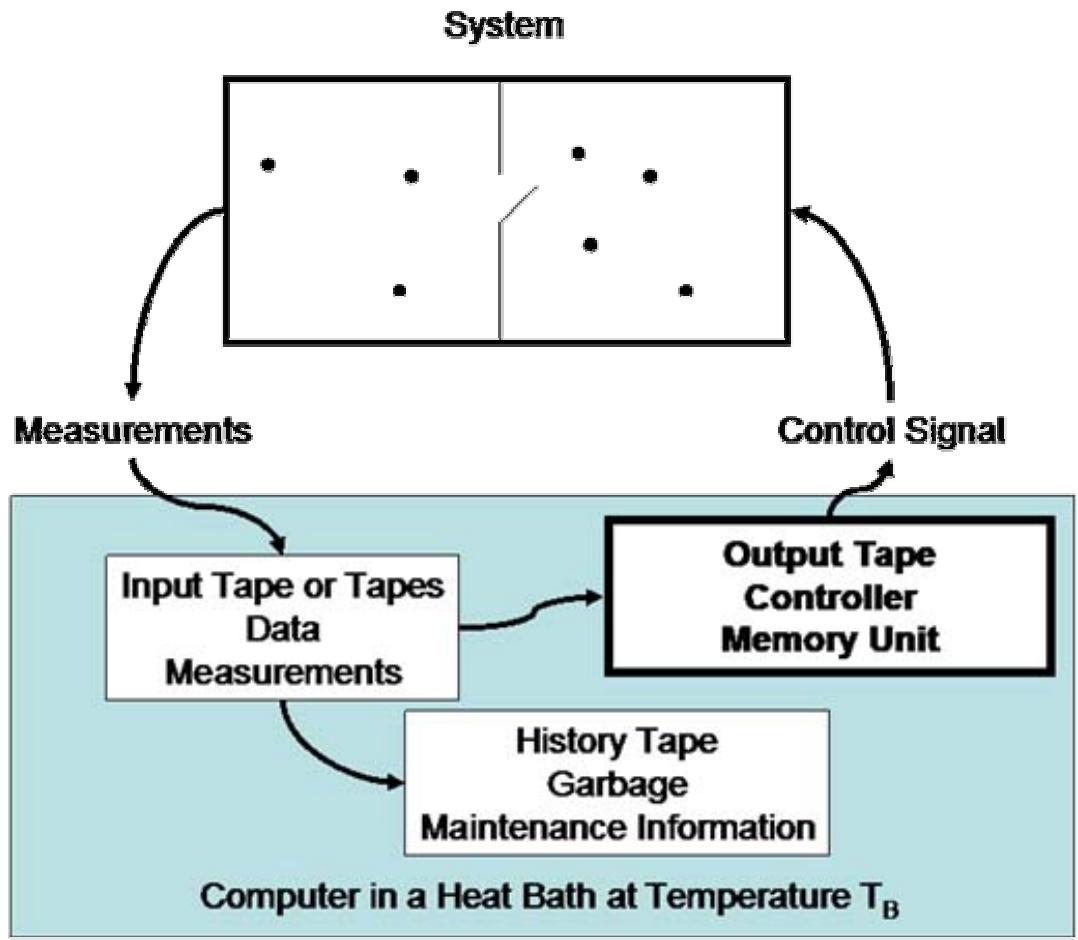

**Figure 1: The Maxwell Demon observes the particles and writes relevant information to an Input Tape (or Tapes) that is immersed in a Heat Bath. The Input Tape is also known as "Data" or "Measurements." The Computer extracts relevant control information from the Input Tape and writes the information to an Output Tape, or "Controller," or "Memory Unit," that sends control signals to a barrier that allows particles to pass or not. The remaining nonessential information is written to a History Tape, or "Garbage Tape," or**

**Complexica Report 031128**

**"Maintenance Tape." The communication channel to the computer is the Measurement Channel. The communication channel from the computer is the Control Channel.**

Our version of the problem is illustrated in Figure 1. We imagine that we have a box of $N$ particles identical in every way except that half of the particles are labeled $A$ and half are labeled $B$. There is a barrier that divides the box in half. As in the original problem there is a trapdoor that, when closed, allows particles to access only the right, $R$, or left, $L$, side of the box and, when open, allows the particles to access the entire box. The trapdoor can be opened or closed in zero time and it absorbs no energy in the process of opening or closing. When a particle passes through the trapdoor it mixes with the particles on its side of the box infinitely fast so that the probability of a particle of a particular type on a particular side approaching the trapdoor is simply proportional to the number of particles of that type on that side. We measure time by counting the number of particles that approach the trapdoor Time $t$, therefore, is the time associated with the $t^{th}$ particle approaching the trapdoor.

The Demon in this paper is an idealized computer immersed in a Heat Bath at temperature $T_B$. The Demon's goal is to sort the particles so that all $A$ particles are on the left and all $B$ particles are on the right. To do this it observes the particles in the box and controls the trapdoor. There is a communication channel, the Measurement Channel, to the computer from the System of particles that transmits measurement information and one, the Control Channel, from the computer to the System that transmits control information. We assume that the channels are lossless. No information is lost and no energy is consumed in the transmission. We neglect any energy transfer or entropy increase in the measurement process. We do, however, consider the entropy increase of the computer and the Heat Bath as it writes to and erases information from the computer.

The computer itself is composed of a number of tapes. The tapes have been given various names by various authors. We will use several of the names in this paper depending on the context of the discussion. The most important tape is labeled "Output Tape" or "Controller" or "Memory Unit." At each time step, the Memory Unit contains information on whether the trapdoor should be open or closed at time $t$. The Memory Unit transmits this information to the System causing the trapdoor to take the appropriate action. The writing and erasing of the Memory Unit may be logically reversible or irreversible depending on the existence of information on the other tapes. As Landauer pointed out,[8] irreversible writing and erasure of a memory unit transfers an amount of heat of approximately $k_B T_B$ per bit to the Heat Bath surrounding the unit. Here $k_B$ is Boltzmann's constant. If the erasure is reversible, then no heat is transferred to the Heat Bath and the Heat Bath does not increase in entropy. Since the Memory Unit returns to its original state after erasure, the entropy change of the Memory Unit is zero.

If the measurement process is optimal, then the Memory Unit is the only tape needed to control the System.[4] Measurement information flows directly from the System to the Memory Unit without passing through the intermediary of other tapes. Here, "optimal" means that exactly the same amount of information is flowing in the measurement transmission line as in the control line. This can occur for instance if the measurement process asks the question, "Is the particle that is approaching the trapdoor on the wrong side of the box or on the correct side of the box?" The answer to this question is completely correlated with whether the Memory Unit sends an open or a closed control signal to the trapdoor. If particles are initially uniformly distributed throughout the box, then the measurement process initially collects one bit of information per measurement and the Memory Unit transmits one bit of information per measurement.

If the measurement process, however, collects more information than is needed by the Memory Unit to control the trapdoor, then a mechanism is needed to condense the information. This can occur, for instance, if the measurement process asks the questions, "Is the particle that is approaching the trapdoor a particle of type $A$ or a particle of type $B$?" … AND … "Is the particle approaching from the right or from the left?" Assuming an initially uniform distribution of particles, the measurement process initially collects two bits of information per measurement while the Memory Unit still transmits only one bit per measurement.

Excess information is stripped from the signal to the Output Tape. The unnecessary excess information is stored on the History Tape or "Garbage Tape" or "Maintenance Information Tape" and the necessary information is stored on the Output Tape. No information is destroyed. The Maintenance Tape and the Output Tape together contain exactly the same information as the Input Tape. Bennett pointed out[9] that, in a logically reversible situation such as this, the Output, and Maintenance Tapes can be erased reversibly after the sorting has been carried out leaving only the Input Tape. Erasure of these tapes therefore can be performed without any transfer of heat to the Bath and consequently no contribution to the change in entropy of the universe. Heat is transferred to the Heat Bath increasing the entropy of the universe only when the Input Tape is erased. This is a logically irreversible process because once the tape is erased, there is no information remaining on any tape to restore the Input. The information on the Input Tape has been irretrievably lost. The increase in entropy of the Heat Bath is simply the information that was stored on the Input Tape. Conversely, we can reversibly erase the Input Tape leaving the Output Tape and the History Tape. The entropy of the Heat Bath increases when we irreversibly erase the surviving two tapes. The entropy increase of the universe here will be the same as irreversibly erasing the Input Tape. Physical examples of this process have been colorfully described by Feynman.[10]

### 3. Reversible and Irreversible Computation

We can make the discussion in the previous section more concrete. Suppose we are measuring two pieces of information about the particle approaching the trapdoor, the type of particle ($A$ or $B$), and the side from which the particle is approaching ($R$ or $L$). We imagine we have two Input Tapes. We write the first piece of information to the first Input Tape and the second piece of information to the second tape. Now, we imagine that we have two more blank tapes labeled "Output Tape" and "History Tape." The Output Tape can contain the information $C$ or $O$ depending on whether the trapdoor should be closed or open. The History Tape can contain the same information as the second Input

Tape, $R$ or $L$. We write from the Input Tapes to the Output and History Tapes according to the following instruction set.

$$AL \to CL$$

$$AR \to OR$$

$$BL \to OL$$ (3.1)

$$BR \to CR$$

The set is interpreted as "If an $A$ particle appears on the left, then write $C$ to the Output Tape and $L$ to the History Tape. If an $A$ particle appears on the right, then write $O$ to the Output Tape and $R$ to the History Tape. Etc." This instruction set directs $A$ particles to the left side of the box and $B$ particles to the right.

This instruction set is reversible in that if we know, for instance, that the trapdoor signal is $C$ and the particle is approaching from the left, then we know that the particle is of type $A$. Therefore, if we know the contents of the Input Tape we can construct the Output and History Tapes. If we know the contents of the Output and History Tapes, we can reconstruct the Input Tape. Then, from Bennett's arguments, we can reversibly erase either the tapes associated with the left side of Eq. (3.1) or with the right side of the equation. There need be no heat transfer to the Heat Bath in the process. Heat transfer occurs when we erase the surviving tape(s). The heat transfer is the same irrespective of whether the surviving tape is the Input Tape or are the Output and History Tapes.

From Zurek's arguments[5] the total entropy increase of the universe due to the sorting is the sum of the physical decrease in entropy of the sorted particles plus the number of bits in the surviving tapes after all reversible erasures have taken place plus the negligible number of bits in the instruction set, Eq. (3.1).

The argument still holds if we make measurements in a different, more efficient, way. Suppose that instead of measuring the type and side of the approaching particle, we just measure whether the approaching particle is on the wrong side or on the proper side. We only need one tape to do this. In that case, the Input Tape has exactly the same amount of information as the Output Tape. No information is written to the History Tape. The instruction set becomes

$$W \to O$$

$$P \to C$$ (3.2)

which can be interpreted as "If a particle is on the wrong side of the box (the Input Tape reads $W$), write $O$ to the Output Tape. If the particle is on the proper side of the box (the Input Tape reads $P$), write $C$ to the Output Tape." Once again the computation is reversible and either the Input or the Output Tape can be erased with no entropy increase to the universe. The entropy of the universe increases when the final surviving tape is erased. The amount of information on the tapes is smaller in this case than in the case in which two pieces of information are measured. Therefore, the increase of the entropy of the universe is smaller in this situation. If the amount of information in the Input Tape equals the amount in the Output Tape, then the design of the measurement process is in some sense optimal. This is the situation discussed in Ref. 4 if we reversibly erase the Input Tape after each measurement.

## 4. Entropy Increase of the Universe

We now have the background we need to explicitly calculate the entropy increase of the universe due to the sorting process. We are interested in two initial conditions:
- *Unsorted to Sorted* - The particles are randomly scattered throughout the box. Thus we are going from an unsorted state to a final state in which $A$ particles are on the left and $B$ particles are on the right – a sorted state.
- *Anti-Sorted to Sorted* - The particles start out in the opposite state to the final state, $A$ particles on the right and $B$ particles on the left. Here we are going from a sorted state to another, but different, sorted state – a resorted state.

In each of the cases we assume that the measurements collect information on what type of particle is approaching the trapdoor as well as from which side the particle is approaching.

### a. From Unsorted to Sorted

#### i. Probabilities

Initially the particles are spread uniformly throughout the box. Half the particles are type $A$ and half are type $B$. Therefore, at any instant, we expect that there are N/4 $A$ particles on the left side of he box, N/4 $A$ particles on the right side of the box, N/4 $B$ particles on the left side and N/4 $B$ particles on the right side. The trapdoor is open, so the particles mix freely. The probability of reaching anywhere in the box and selecting an $A$ particle is ½. The same is true for $B$ particles. Now imagine that a particle is approaching the trapdoor. The Demon decides whether to keep the barrier open or to close it based on the information about the particle he has just collected. If the barrier remains open, then a particle passes from one side to the other. If the particle approaches from the left then the number of particles on the right is increased by one and the number on the left decreases by one and *vice versa* if the particle approaches from the right. If the barrier closes upon the approach of the particle, then the number on each side remains unchanged. Initially, at time 0, we expect an $A$ particle to pass through the barrier a quarter of the time. Therefore, at time 1, the expected the expected number of $A$ particles on the left is

$$\frac{N}{4} + \frac{n(A,R)}{N} = \frac{N}{4} + \frac{1}{4}$$

where $n(A,R)$ is the number of $A$ particles on the right at time zero. The expected number on the right is

$$\frac{N}{4} - \frac{n(A,R)}{N} = \frac{N}{4} - \frac{1}{4}.$$

The same arguments hold for increasing/decreasing the number of $B$ particles on the right/left.

$$\frac{N}{4} + \frac{n(B,R)}{N} = \frac{N}{4} + \frac{1}{4}$$

$$\frac{N}{4} - \frac{n(B,L)}{N} = \frac{N}{4} - \frac{1}{4}$$

The probability of finding a particle $A$ on the right is

$$p_t(A,R) = \frac{n_t(A,R)}{N}. \tag{4.1}$$

where the subscript $t$ indicates the quantity is measured at time $t$. The probabilities for the other three situations can be expressed similarly.

The time dependence of the probabilities can be solved (Appendix A) to yield

$$\begin{aligned} p_t(A,L) &= p_t(B,R) = \frac{1}{2}(1 - \Pi_t) \\ p_t(A,R) &= p_t(B,L) = \frac{1}{2}\Pi_t \end{aligned} \tag{4.2}$$

where

$$\Pi_t \equiv \frac{1}{2}\left(1 - \frac{1}{N}\right)^t.$$

The probability of finding a particle on the wrong side decreases with time and the probability of finding one on the correct side increases.

The solutions, Eq. (4.2), for the probabilities allow us to calculate (Appendix A) the change in entropy as a result of sorting for the System, the Controller, the Input and the Maintenance. We neglect the small amount of information in the instruction sets, Eqs. (3.1) and (3.2).

ii.    System Entropy Change

The initial entropy of the System is

$$S_0 = N. \tag{4.3}$$

Here we are measuring entropy in bits.

After the particles are sorted, the System is in an ordered state with zero entropy

$$S_f = 0 \tag{4.4}$$

The change in System entropy is

$$\Delta S_{system} = -N \ . \tag{4.5}$$

The minus sign indicates that the System has become more ordered through sorting. Equation (4.5) simply states that at least $N$ binary decisions have to be made to sort the particles from an initially random state.

    iii.    Output Tape (Controller) Entropy Change

The change in entropy of the Controller is (Appendix A)

$$\Delta S_{controller} = N + \eta N \tag{4.6}$$

where for large $N$

$$\eta = \frac{\pi^2}{12\ln(2)} - \frac{\ln(2)}{2}$$
$$\approx 0.8399955201358 \tag{4.7}$$

where the logarithm is a natural logarithm.

The Controller entropy can be expressed as

$$\Delta S_{controller} = -\Delta S_{system} + \delta S \ . \tag{4.8}$$

The information is then converted to an entropy increase of the Bath by the same amount as the Memory Unit is erased. This can be expressed as

$$\Delta S_{universe} = \Delta S_{system} + \Delta S_{bath}$$

$$= \Delta S_{system} - \Delta S_{system} + \delta S \tag{4.9}$$

$$= \delta S$$

The $N$ bits of entropy that the System lost by sorting are regained by the Bath. There is additional excess entropy that is gained by the Bath of

$$\delta S = \eta N. \tag{4.10}$$

This represents an excess level of disorder in the Bath over that of the System. It is a "cost" of sorting.

The constant $\eta$ is fundamental. In reality, $\eta$ is not a constant, but has some weak $N$ dependence explicitly given by

$$\eta = \sum_{t=0}^{\infty} \left\{ -\frac{1}{N} \Pi_t \lg[\Pi_t] - \frac{1}{N}(1-\Pi_t) \lg[1-\Pi_t] \right\} - 1 \tag{4.11}$$

and displayed in Figure 2. The logarithm in Eq. (4.11) is to the base 2. For $N > 1000$, Eq. (4.7) is a very good approximation to Eq. (4.11), in other words, in the limit of large N, $\eta$ approaches the constant value of Eq. (4.7).

    iv.    Entropy Change of the Input Tape

The entropy change of the Input Tape can be written (Appendix A)

$$\Delta S_{input} = \Delta S_{controller} + I_M \tag{4.12}$$

where

$$I_M \equiv \sum_{t=0}^{\infty} 1 \tag{4.13}$$

the *Maintenance Information* is the excess information over the Controller requirements collected by the Demon. The controller information is stored on the Output Tape. The Maintenance Information is stored on the Maintenance Tape.

    v.    Entropy Change of the Universe

The information from the surviving tapes is converted to an entropy increase of the Bath by the same amount as the number of bits stored on the surviving tapes. This can be expressed as

$$\Delta S_{universe} = \Delta S_{system} + \Delta S_{bath}$$

$$= \Delta S_{system} + \Delta S_{input}$$

$$= \Delta S_{system} + \Delta S_{controller} + I_M \qquad (4.14)$$

$$= \Delta S_{system} - \Delta S_{system} + \delta S + I_M$$

$$= \delta S + I_M$$

The $N$ bits of entropy that the System lost by sorting are regained by the Bath. There is additional excess entropy that is gained by the Bath of

$$\delta S = \eta N + I_M. \qquad (4.15)$$

This represents an excess level of disorder in the Bath over that of the System. It is a "cost" of sorting. If the measurements are optimal, then the Measurement Information is zero and the entropy increase of the universe is

$$\delta S = \eta N. \qquad (4.16)$$

### b. From Anti-Sorted to Sorted

We consider the initial state in which all the $A$ particles are on the right and all the $B$ particles are on the left. The job of the Maxwell Demon is to move the $A$ particles to the left side of the box and the $B$ particles to the right.

      i.    Probabilities

In this case the probabilities are given by (Appendix B)

$$p_t(A,L) = p_t(B,R) = \left(\frac{1}{2} - \Pi_t\right)$$

$$p_t(A,R) = p_t(B,L) = \Pi_t \qquad (4.17)$$

where again

$$\Pi_t \equiv \frac{1}{2}\left(1 - \frac{1}{N}\right)^t.$$

      ii.    System Entropy Change

Since the System is transitioning from an ordered state to a similar ordered state, the entropy change of the System is zero.

$$\Delta S^*_{system} = 0 \qquad (4.18)$$

      iii.    Entropy of the Controller

The entropy of the Controller is (Appendix B)

$$\Delta S^*_{controller} = \eta^* N$$

$$\equiv \delta S^* \qquad (4.19)$$

where $\eta^* \approx 2.373138220832 \text{ bits}$.

There is some weak $N$ dependence of $\eta^*$ given by

$$\eta^* = \sum_{t=0}^{\infty} \left\{ -\frac{1}{N} 2\Pi_t \lg[2\Pi_t] - \frac{1}{N}(1 - 2\Pi_t)\lg[1 - 2\Pi_t] \right\}. \qquad (4.20)$$

The dependence is displayed in Figure 3.

      iv.    Entropy of the Input Tape

The entropy of the information stored on the Input Tape is (Appendix B)

$$\Delta S^*_{input} = \Delta S^*_{controller} + I_M . \qquad (4.21)$$

We see that, once again, the Demon collects one bit per time step as *Maintenance Information.*

      v.    Entropy Change of the Universe

The information from the surviving tapes is converted to an entropy increase of the Bath by the same amount as the number of bits stored on the surviving tapes. This can be expressed as

$$\Delta S^*_{universe} = \Delta S^*_{system} + \Delta S^*_{bath}$$

$$= \Delta S^*_{input}$$

$$= \Delta S^*_{controller} + I_M \quad (4.22)$$

$$= \delta S^* + I_M$$

The $N$ bits of entropy that the System lost by sorting are regained by the Bath. There is additional excess entropy that is gained by the Bath of

$$\delta S^* = \eta^* N + I_M . \quad (4.23)$$

Once again, if the measurements are optimal, then the Measurement Information is zero and the entropy increase of the universe is

$$\delta S^* = \eta^* N . \quad (4.24)$$

## *5. Concluding Remarks*

In this paper we examined the information flows and entropy generation in the Maxwell Demon problem. We have attempted to treat the Maxwell Demon problem as purely a problem in computation, making the connection to physics through general principles developed by Landauer, Bennett, and Zurek (Refs. 5, 8, and 9). This approach complements an earlier treatment of the problem that takes a purely physical point-of-view.[4] The separation of the computational and information aspects of the Maxwell Demon problem from physics will allow us to relate nonphysical systems to thermodynamics.[6]

We found that, in a physics-free optimally-measured generalization of the problem, the entropy of the universe increased by

$$\delta S = \eta N \text{ bits}$$

in going from a uniform distribution of particles to sorted particles and

$$\delta S^* = \eta^* N \text{ bits}$$

in going from anti-sorted particles to sorted particles. Here, $N$ is the number of particles sorted and $\eta$ and $\eta^*$ are given by Eqs. (1.2) and (1.4).

**Complexica Report 031128**

The entropy accounting is nontrivial. The various components in the problem transmit information or change entropy in non-obvious ways. For instance, the information flow and entropy generation for the unsorted to sorted case is

| | |
|---|---|
| Change in entropy of the particles in the box (order creation) | $-N$ |
| Total information transmitted in measurement channel | $(1+\eta)N + I_M$ |
| Entropy change of the tapes after erasure | $0$ |
| Information written to Input Tape (Data, Measurement) | $(1+\eta)N + I_M$ |
| Information written to Output Tape (Controller, Memory Unit) | $(1+\eta)N$ |
| Information written to History Tape (Garbage, Maintenance) | $I_M$ |
| Entropy change of the heat bath (heat generation) | $(1+\eta)N + I_M$ |
| Total information transmitted in control channel | $(1+\eta)N$ |
| Total entropy change of the universe (particles + bath) | $\eta N + I_M$ |

In the case that the measurements are optimal, the Maintenance Information, $I_M$, is zero.

Even in the case in which there is no net entropy change of the particles there is heat generation and the entropy of the universe increases. Specifically, the information flow and entropy generation for the anti-sorted to sorted case is

| | |
|---|---|
| Change in entropy of the particles in the box (order creation) | $0$ |
| Total information transmitted in measurement channel | $\eta^* N + I_M$ |
| Entropy change of the tapes after erasure | $0$ |
| Information written to Input Tape (Data, Measurement) | $\eta^* N + I_M$ |
| Information written to Output Tape (Controller, Memory Unit) | $\eta^* N$ |
| Information written to History Tape (Garbage, Maintenance) | $I_M$ |
| Entropy change of the heat bath (heat generation) | $\eta^* N + I_M$ |
| Total information transmitted in control channel | $\eta^* N$ |
| Total entropy change of the universe (particles + bath) | $\eta^* N + I_M$ |

Again, in the case that the measurements are optimal, the Maintenance Information, $I_M$, is zero.

There are two sources of "waste" in the problem. The first source of waste is associated with the Maintenance Information. This is excess information that is collected in measurements that is not used to control the System. This excess information ends up as irreversible heat generation. The second source of waste is due to the fact that a particle may approach the trapdoor more than once. This requires multiple measurements on a single particle. This waste is associated with the constants $\eta$ and $\eta^*$. An amount of heat proportional to these constants is irreversibly generated. The total entropy increase of the universe is due to the sum of these two waste terms.

There is another source of heat in the problem. This is heat generation that compensates for the net ordering of the particles. The entropy associated with this heat is fully balanced by the entropy decrease of the sorted particles.

One can calculate the efficiency of an engine driven by the Maxwell Demon sorting process. This is the topic of a future paper.

## *Appendix A: Entropy Changes for the Sorting Problem – Unsorted to Sorted*

### The Probabilities

Initially, particles are distributed randomly throughout the box. After sorting, all *A* particles are on the left and *B* particles are on the right. At time $t = 0$ the numbers of particles in four states are

$$n(A, L) = n(A, R) = n(B, L) = n(B, R) = \frac{N}{4} \qquad (4.25)$$

where

*N* is the Total number of particles

*A* denotes particles of type *A*

*B* denotes particles of type *B*

*L* identifies the left side of the box

*R* identifies the right side of the box so that $n(A, L)$, for instance, is the number of particles of type *A* in the left hand side of the box. When the Controller takes action (i.e. the trapdoor opens or closes) the numbers of particles on each side are either unchanged or incremented/decremented by one. If a particle approaching the trapdoor is on the wrong side, an *A* particle on the right side or a *B* particle on the left side, the trapdoor is opened and allowed to pass through increasing the number of particles on the "correct" side by one and decreasing the number of particles on the "incorrect" side by one. If an approaching particle is on the correct side, the trapdoor remains closed and the numbers of particles on either side remain unchanged. Therefore, the expected number of particles passing from right to left, when a single particle approaches the trapdoor is

$$p(A, R) = \frac{n(A, R)}{N} \leq 1$$

and the expected number passing from left to right is

$$p(B, L) = \frac{n(B, L)}{N} \leq 1.$$

This can be written

$$\begin{aligned} n_{t+1}(A, L) &= n_t(A, L) + p_t(A, R) \\ n_{t+1}(A, R) &= n_t(A, R) - p_t(A, R) \\ n_{t+1}(B, L) &= n_t(B, L) - p_t(B, L) \\ n_{t+1}(B, R) &= n_t(B, R) + p_t(B, L) \end{aligned} \qquad (4.26)$$

where the subscript *t* indicates the $t^{th}$ particle that approached the barrier.

The probabilities for the four classes of particles become

$$p_{t+1}(A,L) = p_t(A,L) + \frac{1}{N}p_t(A,R)$$

$$p_{t+1}(A,R) = p_t(A,R) - \frac{1}{N}p_t(A,R)$$

$$p_{t+1}(B,L) = p_t(B,L) - \frac{1}{N}p_t(B,L)$$

$$p_{t+1}(B,R) = p_t(B,R) + \frac{1}{N}p_t(B,L)$$

(4.27)

Note that

$$p_{t+1}(A,L) + p_{t+1}(A,R) = p_t(A,L) + p_t(A,R) = p_0(A,L) + p_0(A,R) = \frac{1}{2}$$

$$p_{t+1}(B,L) + p_{t+1}(B,R) = p_t(B,L) + p_t(B,R) = p_0(B,L) + p_0(B,R) = \frac{1}{2}$$

Using (4.25), (4.27) can be solved to yield

$$p_t(A,L) = \frac{1}{2} - \frac{1}{4}\left(1 - \frac{1}{N}\right)^t = \frac{1}{2}(1 - \Pi_t)$$

$$p_t(A,R) = \frac{1}{4}\left(1 - \frac{1}{N}\right)^t = \frac{1}{2}\Pi_t$$

$$p_t(B,L) = \frac{1}{4}\left(1 - \frac{1}{N}\right)^t$$

$$p_t(B,R) = \frac{1}{2} - \frac{1}{4}\left(1 - \frac{1}{N}\right)^t$$

(4.28)

where

$$\Pi_t \equiv \frac{1}{2}\left(1 - \frac{1}{N}\right)^t.$$

## Change in Entropy of the System

The entropy of the System before the sorting is

$$\frac{S_0}{N} = -[p_0(A,L) + p_0(A,R)]\lg[p_0(A,L) + p_0(A,R)]$$
$$\quad - [p_0(B,L) + p_0(B,R)]\lg[p_0(B,L) + p_0(B,R)]$$
$$= -2\left[\frac{1}{2}\lg\left(\frac{1}{2}\right)\right]$$
$$= 1$$

(4.29)

After the sorting the probabilities are

$$p_\infty(A,L) = p_\infty(B,R) = 1$$
$$p_\infty(A,R) = p_\infty(B,L) = 0$$

and the entropy is

$$\frac{S_\infty}{N} = -[p_\infty(A,L)]\lg[p_\infty(A,L)] - [p_\infty(B,R)]\lg[p_\infty(B,R)]$$
$$= -2[\lg(1)] \qquad (4.30)$$
$$= 0$$

yielding a change in entropy for the System of

$$\Delta S_{system} = -N. \qquad (4.31)$$

## Change in Entropy of the Controller

The entropy of the Controller is defined as the entropy of the time series of actions of the Controller. The time series can be displayed, for example, as

$$[c,o,c,c,o,c,o,o,o,c,\ldots] \qquad (4.32)$$

where $c$ represents a closed barrier and $o$ represents an open barrier.

The probability that the trapdoor is open/closed is

$$p_t(o) = p_t(A,R) + p_t(B,L) = \Pi_t$$
$$p_t(c) = 1 - p_t(o) = 1 - \Pi_t \qquad (4.33)$$

Before the sorting has started the trapdoor is always open and the entropy of the Controller is zero. After the sorting is complete, the entropy of the Controller is

$$S_\infty = \sum_{t=0}^{\infty}\{-\Pi_t\lg[\Pi_t] - (1-\Pi_t)\lg[1-\Pi_t]\}. \qquad (4.34)$$

Equation (4.34) can be evaluated numerically to yield

$$\Delta S_{controller} = N + \eta N$$

where for large N we have $\qquad (4.35)$

$$\eta \approx 0.83999551.$$

The quantity $\eta$ (eta) is fundamental. We note that $\eta$ has some functional dependence on $N$. Explicitly, $\eta$ is

$$\eta = \sum_{t=0}^{\infty}\left\{-\frac{1}{N}\Pi_t\lg[\Pi_t] - \frac{1}{N}(1-\Pi_t)\lg[1-\Pi_t]\right\} - 1 \qquad (4.36)$$

The dependence is displayed in Figure 2.

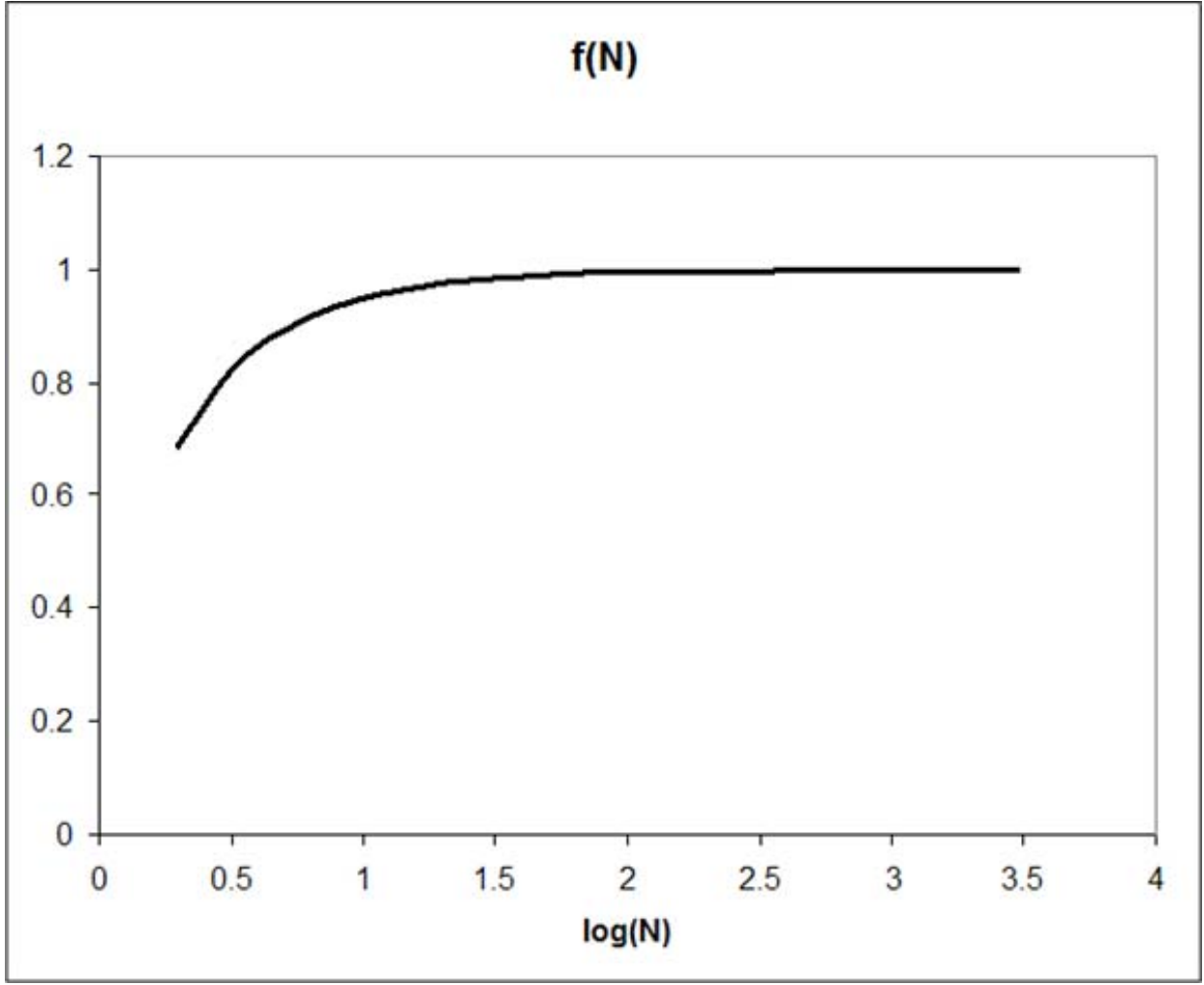

**Figure 2: Functional dependence of** $\eta$ **on** $N$ . **The plot is of** $f(N) \equiv \eta/0.83999551$ . **The log is to the base 10.**

The quantity

$$\delta S \equiv \eta N \qquad (4.37)$$

is the *Unsorted Excess Entropy* required by the Controller to sort the particles. Equation (4.35) can be rewritten as

$$\Delta S_{\text{controller}} = -\Delta S_{\text{system}} + \delta S \qquad (4.38)$$

where $\Delta S_{\text{system}} = -N$ is the entropy lost by the System (and consequently gained by the Controller) due to the ordering of the particles and $\delta S = \eta N$ is the extra entropy "cost" to the Controller for ordering the System.

## Entropy of the Input Tape

The role of the Demon is to send the control signal to the Controller. To do this he needs to acquire information from the particles. As a particle approaches the trapdoor, the Demon initially collects two bits of information from the particle, its type and from which side of the barrier it is approaching. As the sorting proceeds, the amount of information collected by the Demon on each particle decreases. By the time the sorting is complete, the Demon knows which type of particle is approaching by knowing from which side the particle is approaching. Therefore after the sorting is complete, the Demon collects only one bit of information from each particle approaching the trapdoor.

The Demon sends a compressed control signal to the trapdoor. Initially, the Demon sends a single bit per particle to the Controller, i.e. whether to open or close the barrier for the particle. After the sorting has occurred, the barrier is always closed and needs no control signal.

The information collected by the Demon (or more accurately, the change in the Input Tape entropy required to store the information collected from the particles) is

$$I = \sum_{t=0}^{\infty}\{-p_t(A,L)\lg[p_t(A,L)]\}$$
$$+ \sum_{t=0}^{\infty}\{-p_t(A,R)\lg[p_t(A,R)]\}$$
$$+ \sum_{t=0}^{\infty}\{-p_t(B,L)\lg[p_t(B,L)]\}$$
$$+ \sum_{t=0}^{\infty}\{-p_t(B,R)\lg[p_t(B,R)]\}$$

$$= \sum_{t=0}^{\infty}\left\{-\Pi_t \lg\left[\frac{1}{2}\Pi_t\right] - (1-\Pi_t)\lg\left[\frac{1}{2}(1-\Pi_t)\right]\right\}$$

$$= -\Delta S_{\text{system}} + \delta S - \sum_{t=0}^{\infty}\lg\left(\frac{1}{2}\right)$$

$$= -\Delta S_{\text{system}} + \delta S + \sum_{t=0}^{\infty} 1 \qquad (4.39)$$

The quantity

$$I_M \equiv \sum_{t=0}^{\infty} 1 \qquad (4.40)$$

is the excess amount of information (1 bit per approach) that is collected by the Demon but is not passed on to the Controller. We call $I_M$ the *Maintenance Information*. This is an additional "cost" that the Demon pays to collect information. The information collected by the Demon can be written

$$I = \Delta S_{controller} + I_M$$
$$= -\Delta S_{system} + \delta S + I_M$$
(4.41)

## Closed Form Solution

In this section extensive use is made of the tables compiled by Gradshteyn and Ryzhik.[11]

$$1 + \eta = S_1 + S_2$$
where
$$S_1 = -\frac{1}{N} \sum_{t=0}^{\infty} \Pi_t \lg(\Pi_t)$$
$$S_2 = -\frac{1}{N} \sum_{t=0}^{\infty} (1-\Pi_t) \lg(1-\Pi_t)$$
$$\Pi_t = \frac{1}{2}\left(1-\frac{1}{N}\right)^t$$
(4.42)

First Term:

$$S_1 = -\frac{1}{N} \sum_{t=0}^{\infty} \frac{1}{2}\left(1-\frac{1}{N}\right)^t \lg\left[\frac{1}{2}\left(1-\frac{1}{N}\right)^t\right]$$

$$= -\frac{1}{N} \sum_{t=0}^{\infty} \frac{1}{2}\left(1-\frac{1}{N}\right)^t \left\{\lg\left(\frac{1}{2}\right) + \lg\left[\left(1-\frac{1}{N}\right)^t\right]\right\}$$

$$= \frac{1}{N} \sum_{t=0}^{\infty} \frac{1}{2}\left(1-\frac{1}{N}\right)^t \left\{1 - t\lg\left[\left(1-\frac{1}{N}\right)\right]\right\}$$

$$= \frac{1}{2N} \sum_{t=0}^{\infty} x^t - \frac{1}{2N} \lg\left[\left(1-\frac{1}{N}\right)\right] \sum_{t=0}^{\infty} tx^t$$
(4.43)

where

$$x = \left(1 - \frac{1}{N}\right)$$

$$S_1 = \frac{1}{2N}\left(\frac{1}{1-x}\right) - \frac{1}{2N}\lg(x)\left[\frac{x}{(1-x)^2}\right]$$

$$S_1 = \frac{1}{2}\left[1 - (N-1)\lg\left(1 - \frac{1}{N}\right)\right]$$

In the limit of large $N$ we have

$$S_1 = \frac{1}{2}\left(1 + \frac{1}{\ln(2)}\right) \approx 1.221347 \tag{4.44}$$

Second Term:

$$S_2 = -\frac{1}{N}\sum_{t=0}^{\infty}\left(1 - \frac{1}{2}x^t\right)\lg\left(1 - \frac{1}{2}x^t\right)$$

$$\tag{4.45}$$

where

$$x = 1 - \frac{1}{N}$$

We use

$$\ln(1-y) = -\sum_{i=1}^{\infty}\frac{y^i}{i} \tag{4.46}$$

to yield

$$S_2 = \frac{1}{2N} + \frac{1}{N}\sum_{t=1}^{\infty}\left\{\frac{\left(1-\frac{1}{2}x^t\right)}{\ln(2)}\sum_{i=1}^{\infty}\frac{\left(\frac{1}{2}x^t\right)^i}{i}\right\}$$

$$= \frac{1}{2N} + \frac{1}{N\ln(2)}\sum_{i=1}^{\infty}\left\{\frac{1}{2^i i}\sum_{i=1}^{\infty}\left(1-\frac{1}{2}x^t\right)x^{it}\right\} \quad (4.47)$$

$$= \frac{1}{2N} + \frac{1}{N\ln(2)}\sum_{i=1}^{\infty}\left\{\frac{1}{2^i i}\left[\left(\frac{x^i}{1-x^i}\right)-\frac{1}{2}\left(\frac{x^{i+1}}{1-x^{i+1}}\right)\right]\right\}$$

which becomes in the large $N$ limit

$$S_2 \approx \frac{1}{2\ln(2)}\sum_{i=1}^{\infty}\left\{\frac{(i+2)}{2^i i^2(i+1)}\right\}$$

$$= \frac{1}{2\ln(2)}\left\{\sum_{i=1}^{\infty}\left\{\frac{2}{2^i i^2}\right\}-\sum_{i=1}^{\infty}\left\{\frac{1}{2^i i}\right\}+\sum_{i=1}^{\infty}\left\{\frac{1}{2^i(i+1)}\right\}\right\} \quad (4.48)$$

$$= \frac{1}{2\ln(2)}\left\{\frac{\pi^2}{6}-\left[\ln(2)\right]^2+\ln(2)-1\right\}$$

which finally yields

$$\eta = S_1 + S_2 = \frac{\pi^2}{12\ln(2)}-\frac{\ln(2)}{2}. \quad (4.49)$$

## *Appendix B: Entropy Changes for the Sorting Problem – Anti-Sorted to Sorted*

### *The Probabilities*

Initially, all $A$ particles are on the right and $B$ particles are on the left. After sorting, all $A$ particles are on the left and $B$ particles are on the right. At time $t = 0$ the numbers of particles in four states are

$$n(A, L) = n(B, R) = 0$$
$$n(B, R) = n(B, L) = \frac{N}{2} \tag{7.1}$$

The probabilities for the four classes of particles are

$$p_{t+1}(A, L) = p_t(A, L) + \frac{1}{N} p_t(A, R)$$
$$p_{t+1}(A, R) = p_t(A, R) - \frac{1}{N} p_t(A, R)$$
$$p_{t+1}(B, L) = p_t(B, L) - \frac{1}{N} p_t(B, L) \tag{7.2}$$
$$p_{t+1}(B, R) = p_t(B, R) + \frac{1}{N} p_t(B, L)$$

The solution is

$$p_t(A, L) = \frac{1}{2} - \frac{1}{2}\left(1 - \frac{1}{N}\right)^t = \left(\frac{1}{2} - \Pi_t\right)$$
$$p_t(A, R) = \frac{1}{2}\left(1 - \frac{1}{N}\right)^t = \Pi_t$$
$$p_t(B, L) = \frac{1}{2}\left(1 - \frac{1}{N}\right)^t = \Pi_t \tag{7.3}$$
$$p_t(A, R) = \frac{1}{2} - \frac{1}{2}\left(1 - \frac{1}{N}\right)^t = \left(\frac{1}{2} - \Pi_t\right)$$

where

$$\Pi_t \equiv \frac{1}{2}\left(1 - \frac{1}{N}\right)^t.$$

### **Change in Entropy of the System**

The entropy of the System before and after the sorting is zero so there is no change in entropy.

$$\Delta S^*_{system} = 0. \tag{7.4}$$

## Change in Entropy of the Controller

The probability that the trapdoor is open/closed is

$$p_t(o) = p_t(A,R) + p_t(B,L) = 2\Pi_t$$
$$p_t(c) = 1 - p_t(o) = 1 - 2\Pi_t \tag{7.5}$$

Before the sorting has started the trapdoor is always open and the entropy of the Controller is zero. After the sorting is complete, the entropy of the Controller is

$$.S_\infty = \sum_{t=0}^{\infty} \{-2\Pi_t \lg[2\Pi_t] - (1-2\Pi_t)\lg[1-2\Pi_t]\} \tag{7.6}$$

Equation (4.34) can be evaluated numerically to yield

$$\Delta S^*_{controller} = \eta^* N$$

where for large N we have $\tag{7.7}$

$$\eta^* \approx 2.37313821$$
.

The quantity $\eta^*$ (eta) is fundamental. We note that $\eta^*$ has some functional dependence on $N$. Explicitly, $\eta^*$ is

$$\eta^* = \sum_{t=0}^{\infty} \left\{ -\frac{1}{N} 2\Pi_t \lg[2\Pi_t] - \frac{1}{N}(1-2\Pi_t)\lg[1-2\Pi_t] \right\} \tag{7.8}$$

The dependence is displayed in Figure 3.

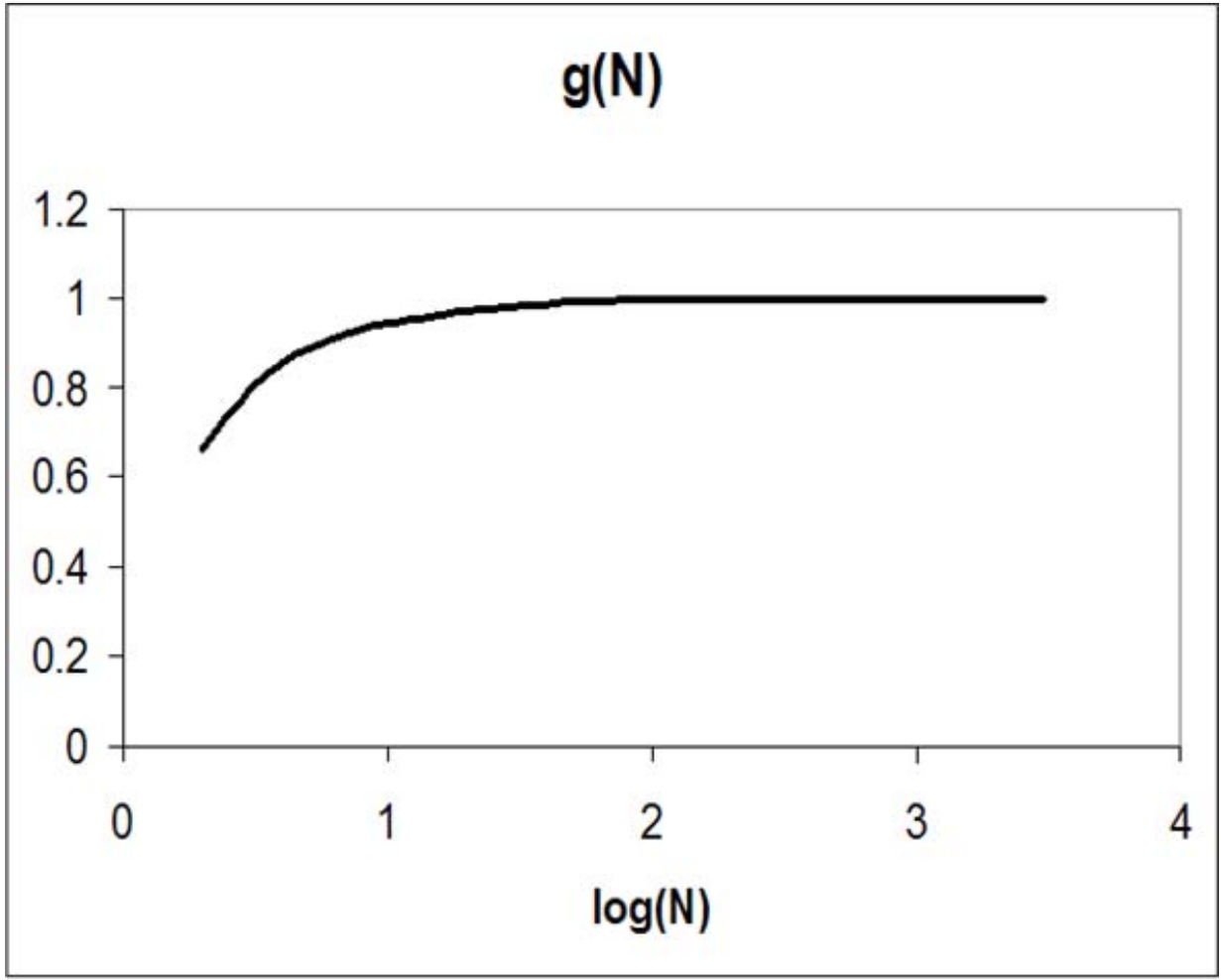

**Figure 3: Functional dependence of** $\eta^*$ **on** $N$ **. The plot is of** $g(N) = \eta^*/2.373138220832$.
**The log is to the base 10.**

The quantity

$$\delta S^* \equiv \eta^* N \qquad (7.9)$$

is the *Excess Entropy* required by the Controller to sort the particles. Equation (4.35) can be rewritten as

$$\Delta S^*_{controller} = \delta S^* \qquad (7.10)$$

where the change in System entropy is zero $\delta S^* = \eta^* N$ is the extra entropy "cost" to the Controller for ordering the System.

## Entropy of the Input Tape

The information collected by the Demon is

$$I^* = \sum_{t=0}^{\infty}\{-p_t(A,L)\lg[p_t(A,L)]\}$$
$$+ \sum_{t=0}^{\infty}\{-p_t(A,R)\lg[p_t(A,R)]\}$$
$$+ \sum_{t=0}^{\infty}\{-p_t(B,L)\lg[p_t(B,L)]\}$$
$$+ \sum_{t=0}^{\infty}\{-p_t(B,R)\lg[p_t(B,R)]\}$$

$$= \sum_{t=0}^{\infty}\left\{-2\Pi_t \lg\left[\frac{1}{2}2\Pi_t\right] - (1-2\Pi_t)\lg\left[\frac{1}{2}(1-2\Pi_t)\right]\right\}$$

$$= \delta S^* + \sum_{t=0}^{\infty} 1$$

$$= \delta S^* + I_m \qquad (7.11)$$

The quantity

$$I_M \equiv \sum_{t=0}^{\infty} 1 \qquad (7.12)$$

is the same *Maintenance Information* we encountered in the maximum entropy case. The information collected by the Demon can be written

$$I^* = \Delta S^*_{\text{controller}} + I_M$$
$$= \delta S^* + I_M \qquad (7.13)$$

## Closed Form Solution

In this section extensive use is made of the tables compiled by Gradshteyn and Ryzhik.[11]

$$\eta^* = S_1 + S_2$$

where

$$S_1 = -\frac{1}{N}\sum_{t=0}^{\infty} 2\Pi_t \lg(2\Pi_t) \qquad (7.14)$$

$$S_2 = -\frac{1}{N}\sum_{t=0}^{\infty}(1-2\Pi_t)\lg(1-2\Pi_t)$$

$$\Pi_t = \frac{1}{2}\left(1-\frac{1}{N}\right)^t$$

First Term:

$$S_1 = -\frac{1}{N}\sum_{t=0}^{\infty}\left(1-\frac{1}{N}\right)^t \lg\left[\left(1-\frac{1}{N}\right)^t\right]$$

$$= \frac{\lg\left[\left(1-\frac{1}{N}\right)\right]}{N}\sum_{t=0}^{\infty} tx^t$$

$$= \frac{1}{\ln(2)} \qquad (7.15)$$

in the large $N$ limit,

where

$$x=\left(1-\frac{1}{N}\right)$$

Second Term:

$$S_2 = -\frac{1}{N}\sum_{t=0}^{\infty}(1-x^t)\lg(1-x^t)$$

$$= \frac{1}{N\ln(2)}\sum_{t=0}^{\infty}(1-x^t)\sum_{i=1}^{\infty}\frac{(x^t)^i}{i}$$

$$= \frac{1}{N\ln(2)}\sum_{i=1}^{\infty}\frac{1}{i}\sum_{t=0}^{\infty}\left[(x^i)^t - (x^{i+1})^t\right]$$

$$= \frac{1}{N\ln(2)}\sum_{i=1}^{\infty}\frac{1}{i}\left(\frac{x^i}{1-x^i} - \frac{x^{i+1}}{1-x^{i+1}}\right)$$

$$= \frac{1}{N\ln(2)}\sum_{i=1}^{\infty}\frac{1}{i}\left(\frac{N}{i} - \frac{N}{i+1}\right)$$

$$= \frac{1}{\ln(2)}\sum_{i=1}^{\infty}\left(\frac{1}{i^2} - \frac{1}{i} + \frac{1}{i+1}\right) \tag{7.16}$$

$$= \frac{1}{\ln(2)}\left(\frac{\pi^2}{6} - 1\right) \text{ in the large } N \text{ limit.}$$

which yields

$$\eta^* = S_1 + S_2 = \frac{\pi^2}{6\ln(2)} \approx 2.373138221. \tag{7.17}$$